\documentclass[aps,prl,twocolumn,amsmath,amssymb]{revtex4}
\usepackage{graphicx}
\usepackage{bm}
\usepackage{color}
\usepackage{psfrag}
\begin{document}
\title{Generic mixed columnar-plaquette 
phases in Rokhsar-Kivelson models}
\author{A. Ralko,${^1}$ 
D. Poilblanc,${^1}$ and R. Moessner${^2}$ }
\affiliation{
${^1}$ Laboratoire de Physique Th\'eorique, CNRS 
and Universit\'e de Toulouse, F-31062 France \\
${^2}$ Rudolf Peierls Centre for Theoretical Physics, Oxford University, 1 Keble Road, 
Oxford OX1 3NP, UK} 
\date{October 5, 2007}
\begin{abstract}
We revisit the phase diagram of Rokhsar-Kivelson models, which 
are used in fields such as superconductivity, frustrated magnetism,
cold bosons, and the physics of Josephson junction arrays. From an
extended height effective theory,
we show that one of two simple generic phase
diagrams contains a mixed phase that interpolates continuously between
columnar and plaquette states. For the square lattice quantum dimer model 
we
present evidence from exact diagonalization and Green's function
Monte Carlo techniques that this scenario is realised, by
combining an analysis of the excitation gaps of
different symmetry sectors 
with information on plaquette structure factors.  This
presents a natural framework for resolving the disagreement between
previous studies.
\end{abstract}
\maketitle

The Rokhsar-Kivelson Quantum Dimer model (RK-QDM)~\cite{rokhsar} on
the square lattice, originally proposed in the context of
high-temperature superconductivity, and its descendants have taken on
a central role in the study of quantum systems incorporating a hard
local constraint. They have thus been prominent in the context of
hardcore bosons hopping \cite{senthil} on
frustrated lattices, Josephson junction arrays~\cite{albuquerque},
frustrated Ising models in a transverse field or with small XY
exchange~\cite{moessner}, gauge theories in unusual sectors
\cite{MSF}, spin orbital models~\cite{mila} and cold
atoms~\cite{buechler}.  Studies of this model and its extensions have
unearthed a wealth of phenomena, including instances of deconfined
quantum criticality and a new route to deconfinement\cite{Cantor}.

\begin{figure}[h]
\includegraphics[width=0.4\textwidth,clip]{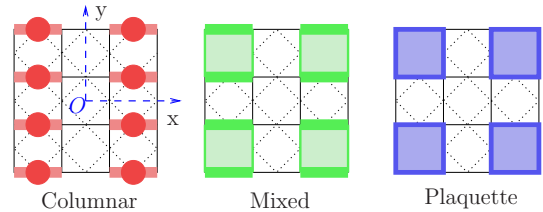}
\caption{\label{phases} Schematic representation of the different 
ordered states considered, the columnar, the mixed columnar-plaquette and the plaquette states. 
Dashed lines show the extra bonds of the checkerboard lattice of the equivalent 
hard-core boson model (see Ref.~\protect\cite{senthil}). On the left, dimers (thick bonds) map
to bosons (red circles). }
\end{figure}

As many RK models belong to the rare class of models of correlated
quantum matter without a sign problem, they are in principle amenable
to efficient numerical study; they also tend to have well-studied
effective field theories formulated in terms of height/gauge
degrees of freedom \cite{MSF}.  
This makes it all the more surprising that there is still
substantial disagreement about the phase structure of the original 
RK
model, the QDM on the square lattice. Indeed, a pioneering study by
Leung et al.\cite{leung} suggested a transition out of a columnar
phase (Fig.~\ref{phases}) to occur as $v/t$ increases from $-\infty$
at $v/t\sim-0.2$ (see Eq.~\ref{eq:hrk}); however, a careful 
detailed recent
investigation by Syljuasen\cite{syljuasen} argued that in fact the
columnar phase persists until $v/t\sim+0.6$

A direct
columnar-plaquette transition {\em en route} to the RK point at 
$v=t$  
is in fact not unusual -- it 
appears to exist in the closely related quantum six-vertex model on
the square \cite{shannon} and the hexagonal QDM~\cite{moessner2}.  
Here, we show that there exists a second {generic} phase
diagram where this first-order transition is replaced by a continuous
interpolation via a mixed phase (Fig.~\ref{phases}). Based on Exact
Diagonalisations (ED) and Green's Function Quantum Monte Carlo (GFMC)
numerics we identify the square lattice QDM as the first candidate for
realising this scenario: around $v/t=0$, we find evidence of a mixed
columnar-plaquette phase, which breaks translational symmetry in two
perpendicular directions like the plaquette, but also $\pi/2$ rotational
symmetry like the columnar phase.  The strongest
evidence for such a phase is provided by a symmetry-based finite-size
scaling analysis of the low energy spectrum. 

The remainder of this paper provides the analysis
backing up these assertions: first we present the pertinent numerical
results on the QDM, followed by a more general analysis of
RK models in $d=2$ whose effective theory can be described by a height
model.

The square lattice QDM Hamiltonian
reads:\cite{rokhsar}
\begin{eqnarray}
\label{eq:hrk}
H = v \sum_{c}  N_{f}(c) |c \rangle \langle c| - t \sum_{(c,c')} |c\rangle
\langle c' | 
\end{eqnarray}
where the sum over $c$ runs on all configurations in the Hilbert space, $N_f
(c)$ is the number of flippable plaquettes contained in $|c\rangle$, {\it
i.e.} number of plaquettes with two parallel dimers, and the sum over $(c,c')$
runs on all configurations $|c\rangle$ and $| c'\rangle$ that differ by a
single dimer flip.

\begin{figure}[h]
\begin{minipage}[c]{1.0\linewidth}
\begin{minipage}[c]{1.0\linewidth}
\hspace{-5.0cm}
\includegraphics[width=0.2\textwidth,clip]{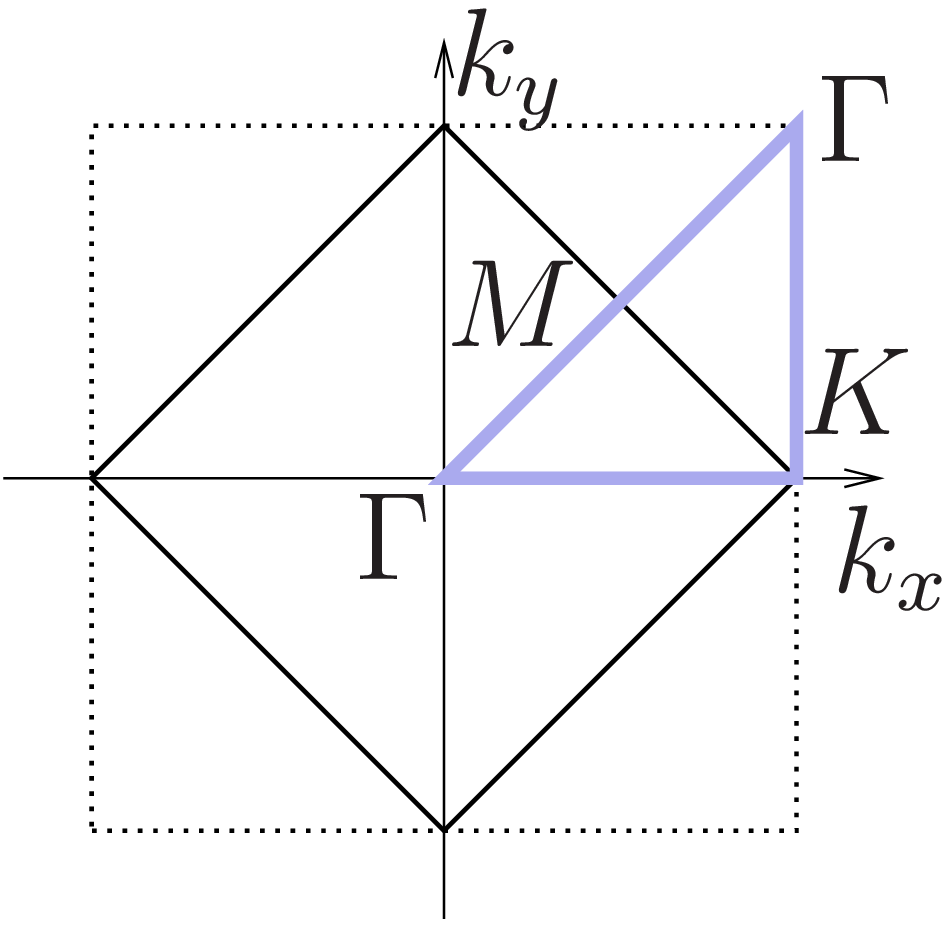}
\vspace{-4cm}

\end{minipage}
\begin{minipage}[c]{1.0\linewidth}
\includegraphics[width=1.0\textwidth,clip]{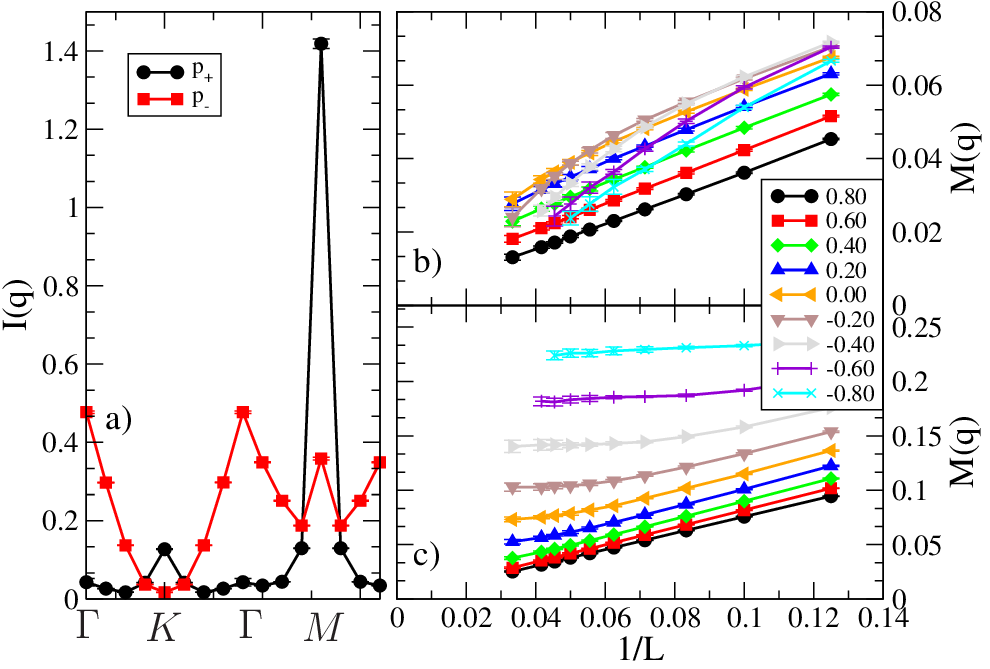}
\end{minipage}
\end{minipage}

\caption{\label{order} a) Behaviour of the structure factor for the 8x8-site
cluster at $v/t=0.2$, along the path of the Brillouin zone depicted in the
inset. A clear peak appears at point $K$. Size-scaling of the order parameters
in the two symmetry sectors are given, for $(K,A_1)$ (b) and $(\Gamma,B_1)$
(c).}
\end{figure}

\emph{Symmetry and first excitation spectrum: }
To compute excitation gaps we combine ED 
with GFMC calculations of dynamical correlations in imaginary time.
In the latter case, the choice of an operator is
crucial in order to ``construct'' an excitation with well-defined
symmetry properties (i.e.\ quantum numbers). 
Starting with the elementary symmetry-breaking patterns of Fig.~\ref{phases},
degenerate ground states with different quantum
numbers can be constructed as listed in Table~\ref{tab_sym}.
On general grounds, symmetry breaking involving any of these states
is signaled on finite clusters of increasing size by the collapse of the 
quasi-degenerate GS and the opening of a robust gap above them to higher
energy excitations. 
Clearly columnar and plaquette phases differ
in two states with different quantum numbers (defined
by a momentum and a point group quantum number),
($\Gamma,B_1$) (point $q=(0,0)$ in the BZ depicted in the inset of
Fig.~\ref{order}, with symmetry $B_1$) for the first one and $(K,A_1)$
($q=(\pi,\pi)$) for the second. 
As shown in \cite{shannon}, a level crossing between two such
energy levels was the signature of a first order phase transition.
In contrast, as we show further down, no such level crossing
is observed here, suggesting a more complex scenario.

\begin{table}[h]
\begin{center}
\begin{ruledtabular}
\begin{tabular}{|*{7}{c|}}
Order&
$\Gamma,A_1$&$\Gamma,B_1$&$M,A_1$&$K,A_1$&$K,B_1$&$M,A_1(*)$\\
\hline
Columnar   &$\surd$&$\surd$&$\surd$  &         &             & \\
Plaquette  &$\surd$&       &$\surd$  &$\surd$  &             & \\
 Mixed     &$\surd$&$\surd$&$\surd$  &$\surd$  &$\surd$&$\surd$ \\
\end{tabular}
\end{ruledtabular}
\end{center}
\caption{\label{tab_sym} Quantum numbers of the degenerate GS characterizing
the ordered phases considered in this paper. We used the standard 
notation of irreducible
representation of the $C_{4v}$ and $C_{2v}$ point group symmetries, acting
around the site $O$ as depicted in Fig.~\ref{phases}. Definition of the
momentum points is given in the Brillouin zone of the inset of
Fig.~\ref{order}. Note that a state at momentum $(0,\pi)$, not shown, is
degenerate with $(\pi,0)$. (*) denotes the first excited level in the
($M,A_1$) sector.}
\end{table}

Since ED are limited to $8\times 8$ clusters
we also used GFMC on clusters of size up to 
$22\times 22$ to calculate gaps in different symmetry sectors
by considering dynamical correlations as defined in [\onlinecite{ralko1}]:
\begin{eqnarray}
D(q,\tau) = \frac{\langle \Psi_G| P_{\alpha}(-q) e^{-H \tau} P_{\alpha}(q)
|\Psi_0 \rangle}{\langle \Psi_G| e^{-H \tau} |\Psi_0 \rangle}
\end{eqnarray}
with $|\Psi_G\rangle$ a guiding function.
To improve the quality of $|\Psi_G\rangle$ we explicitly work in the
boson representation (see Fig.~\ref{phases} and Ref.~\cite{senthil}) which enables to minimize a Jastrow
wave-function with the help of a variational Monte-Carlo method.
$P_{\alpha}(q)$ is a (diagonal) operator with the same symmetry as the
excited state that we intend to target and is defined as the Fourier
transform of plaquette operators
\begin{eqnarray}
\begin{array}{cc}
\includegraphics[width=0.05\textwidth,clip]{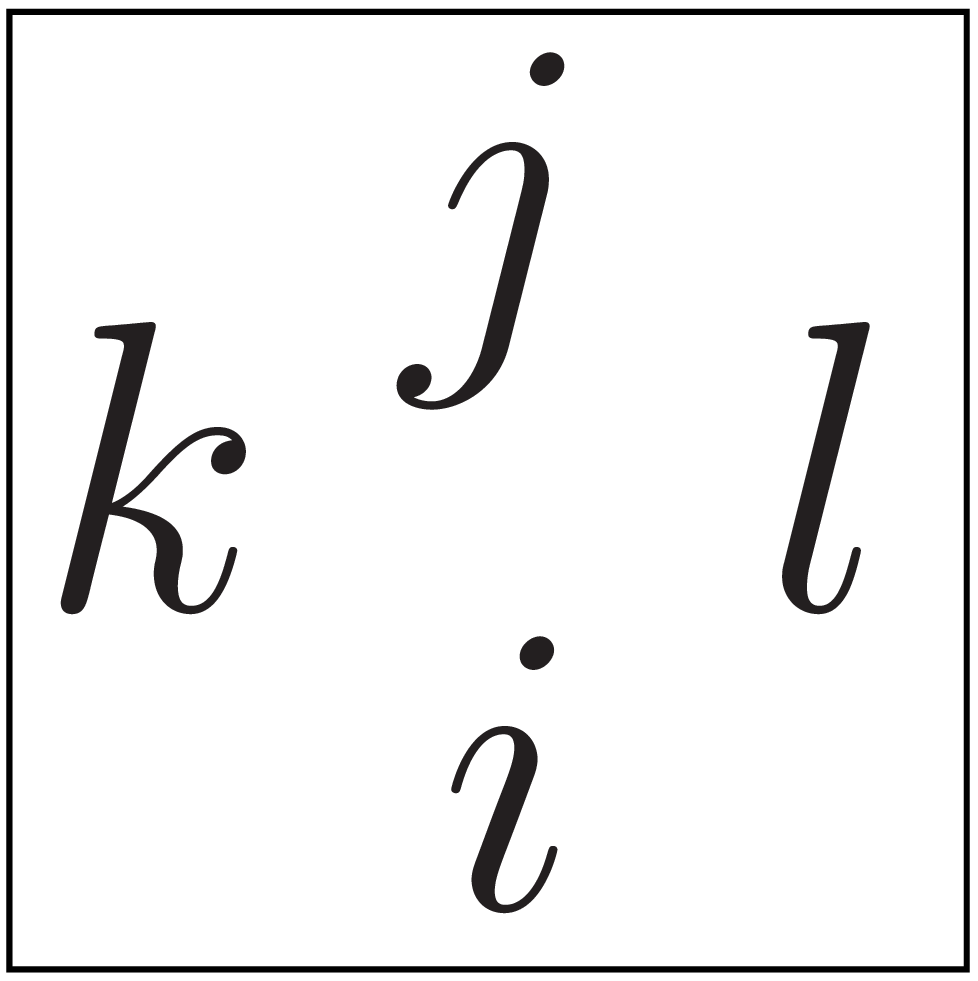}\\
P_{+} = d(i)d(j) + d(k)d(l) \\ 
P_{-} = d(i)d(j) - d(k)d(l) 
\end{array}
\end{eqnarray}
where $d(i)$ is the dimer operator acting on link $i$. 
Note that the point group symmetry of $P_{\alpha}$ depends on the momentum $q$. 
For $P_{+}$, it always corresponds to the most symmetric $A_1$
irreducible representation (IR). 
For $P_{-}$, it corresponds to the $B_1$ IR for the high-symmetry points,
\emph{e.g.} $\Gamma$ and $K$, and to the $A_1$ IR at point $M$.
Note also that the expectation value of $P_{+}(K)$ ($P_{-}(\Gamma)$) is finite
in a generic plaquette (columnar) state.

\begin{figure}[h]
\includegraphics[width=0.45\textwidth,clip]{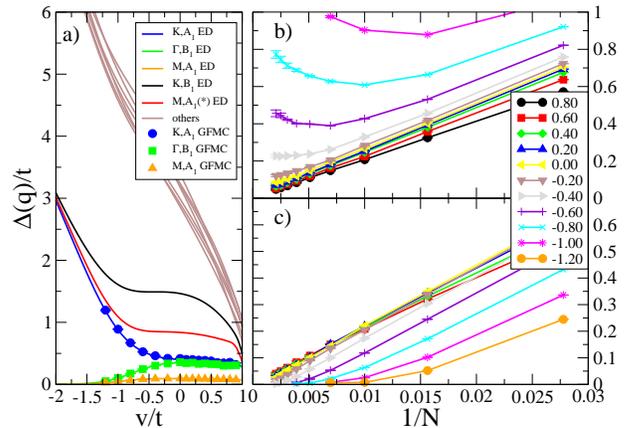}
\caption{\label{gaps} First gap excitations. a) Comparison between ED and
GFMC for the 8x8 site cluster for the interesting q-points. Size-scaling of
the gaps for the $(K,A_1)$ (b) and $(\Gamma,B_1)$ (c) symmetries.}
\end{figure}

Our results for the excitation gaps are
given in Fig.~\ref{gaps}, with q-points defined through the Brillouin zone
displayed in Fig.~\ref{order}.
Panel (a) shows the full exact spectrum on a 8x8-size 
cluster and a comparison with the GFMC estimates for the lowest levels.
We find an excellent agreement which 
validates the GFMC method on larger clusters.
Interestingly, our data distinguish three regimes:
(i) for $v/t\lesssim -0.5$, the $(\Gamma,B_1)$ and $(M,A_1)$ levels 
have very small excitation energy and are well-separated from the rest 
of the spectrum by a sizable gap;
(ii) for roughly $v/t> -0.5$, a second group of states of
$(K,A_1)$, $(M,A_1)$ and $(K,B_1)$ quantum numbers joins 
the previous low energy states
while remaining well separated from other higher energy states; 
(iii) the critical regime near the RK point $v\alt t$.
A comparison with Table~\ref{tab_sym} immediately 
suggests a mixed phase and a columnar phase, respectively, 
for $v/t>-0.5$ and $v/t\lesssim -0.5$.
These data seem already to be inconsistent with
a pure plaquette state for which the $\Delta_{B_1}(\Gamma)$ ($\Delta_{A_1}(K)$) 
gap is finite (vanishing), which would imply a crossing of the $(\Gamma,B_1)$ and 
$(K,A_1)$ levels, in contradiction to what is seen numerically. 
To back up the above, we have performed finite size scalings
of $\Delta_{A_1}(K)$ and $\Delta_{B_1}(\Gamma)$ shown in Fig.~\ref{gaps}(b) and (c),
respectively. 
$\Delta_{A_1}(K)$ clearly extrapolates to zero in the thermodynamic limit 
for $v/t > 0.1$ (compatible with both plaquette and mixed states) 
and opens up near $v/t \simeq 0.0 \pm 0.1$.
Surprisingly,  $\Delta_{B_1}(\Gamma)$ vanishes in the thermodynamic
limit in the whole range of parameters studied (within an accuracy of $10^{-3}$)
with a clear exponential behavior for $v/t \lesssim -0.4$. 
Note that the fact that $\Delta_{B_1}(\Gamma)$ always remains  
smaller than $\Delta_{A_1}(K)$
for all sizes {\it itself} rules out a pure plaquette state.

\emph{Structure factors:} Next we check that the above scenario obtained
from the spectrum analysis (believed to be the most accurate criterium) 
is compatible with the behavior of the related structure factors,
\begin{eqnarray}
I(q) = \frac{\langle \Psi_0 | P_{\alpha} (-q)
P_{\alpha} (q) | \psi_0 \rangle}{\langle \Psi_0 | \Psi_0 \rangle}.
\end{eqnarray}
Results are displayed in Fig.~\ref{order} where the left panel shows
the behaviors of the $P_{+}$ and $P_{-}$ correlations along a given
path of the Brillouin zone (on a 8x8-size cluster) and the right panel
the size-scalings of their respective order parameters $M(q) =
\sqrt{I(q)}/L$, $L$ being the linear size of the system.  These data
correspond to $v/t=0.2$, \emph{i.e.} far from the proposed
plaquette-columnar transition in [\onlinecite{syljuasen}], deep in the
previously-supposed columnar phase.  However, our results 
reveal a Bragg peak at point $K$ which is found to survive upon
extrapolation to the thermodynamic limit.  On the right panel of
Fig.~\ref{order}, size-scalings suggest that columnar order only
develops for $v/t\lesssim0.6$ 
while a finite plaquette order is
present in the range $0.0 \lesssim v/t \lesssim 0.8$.

\begin{figure}[h]
\includegraphics[width=0.50\textwidth,clip]{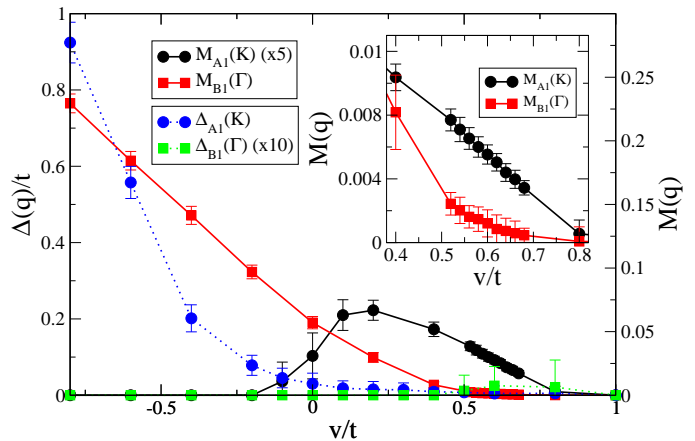}
\caption{\label{tdl} Thermoynamic limit of the order parameters (black and red
curves) and of the first excitation gaps (blue and green dashed curves) in
function of $v/t$. $M_{A_1}(K)$ is multiplied by a factor 5 for clarity. A
zoom around $v/t=0.6$ is shown in the inset. $M_{A_1}(K)$ is not multiplied
here.}
\end{figure}

\emph{Discussion and interpretation of the numerics:} 
Our extrapolated results as a function of $v/t$ for the relevant gap and structure factors 
are summarized in Fig.~\ref{tdl}.
The results for the gaps provide clear evidence for lattice-rotation
symmetry breaking for all parameters ($\Delta_{B_1} (\Gamma)=0$) up to
$v/t=0.6$ (for $v/t>0.6$ a tiny $(\Gamma,B_1)$ gap cannot be
excluded). We find two regimes: (i) For $v/t\lesssim 0.0$,
$\Delta_{A_1} (K)$ is finite, indicating a pure columnar phase; (ii)
for $0.0\lesssim v/t\lesssim 1$, $\Delta_{A_1} (K)$ vanishes,
implying an additional translation-symmetry breaking compatible with
the mixed phase.  The behavior of the structure factors is consistent
with such a picture in this range of parameters.
However, (i) error bars for $-0.1<v/t<0.1$ are still too
large for {\it simultaneous} accurate extrapolations of {\it both}
gaps and order parameters and (ii) upon
approaching the continuous transition at the RK point, both structure
factors become so small (inset in Fig.~\ref{tdl}), that we are
unable to distinguish them from zero within the accuracy of our
simulations for $v/t \agt 0.8$.

Syljuasen argued that rotation symmetry is restored when $v/t>0.6$
signaling the transition to a pure plaquette phase, at least up to
$v/t\simeq 0.9$~\cite{syljuasen}.  This claim 
is consistent
with our results in that a small $(\Gamma,B_1)$ gap cannot be
ruled out for $v/t>0.6$ and that it is plausible that our {\it
diagonal} $P_{+}$ plaquette operator, although of the correct symmetry
to "pick up" plaquette order, only gives a very small weight to the
dynamical correlations when approaching the RK point.  Also, note
that the transition at $v/t = 0.0 \pm 0.1$ that we evidence here from
the columnar state to the novel mixed columnar-plaquette state seems
to be smooth and might be compatible with a second order phase
transition.  We believe our scenario of an intermediate mixed state
reconciles the {\it a priori} conflicting results, (i) the ED data by
Leung \emph{et al.} suggesting the existence of an intermediate phase
for roughly $v/t > -0.2 $~\cite{leung} and (ii) Syljuasen's claim of a
transition at
$v/t=0.6$~\cite{syljuasen}. However, Syljuasen's analysis suggested that
a mixed phase, if present at all, exists in a narrower range of
parameters. At the same time, he pointed out that there was no sign 
of a discontinuous transition. 


\emph{Effective height theory}: We next 
show that the
generic alternative to a first-order plaquette-columnar transition is
a continuous interpolation between them {\em via a mixed phase}, 
in agreement with the scenario obtained above.  
This we do
by reformulating the dimer model in terms of a height variable, $h$,
living on the plaquettes of the square lattice. {\em Mutatis
mutandis}, such a mapping applies to $d=2$ RK models incorporating
constraints which can be cast as local  U(1) conservation laws 
\cite{moessner2,shannon,jspice}. 
 
The mapping from dimers to heights proceeds as follows.
Label the two
sublattices of the square lattice as red and green. When going from
the centre of one plaquette to another in an (anti-)clockwise
direction around a (red) green site, (add) subtract 1 from the value
of the height if no dimer is crossed; and (subtract) add 3 if a dimer
is crossed.  An effective field theory starting from the microscopic
heights can then obtained for a {\em coarse-grained} height field
$h$. This procedure is described in detail in a nice paper by Zeng and
Henley \cite{ZengHenley}, and the corresponding quantum theory is
given in Ref.~\onlinecite{MSF}.

For our purposes, the following facts are important.  Firstly, a `flat
surface', i.e.\ long-range order in $h$, corresponds to a dimer
crystal. In particular, a (half-) integer valued $\langle h\rangle$
corresponds to a (columnar) plaquette phase, whereas intermediate
values of $\langle h\rangle$ correspond to mixed phases like the one
discussed above. Secondly, $h\rightarrow h+1$ amounts to a $\pi/2$
rotation of the dimer configuration, so that $h\rightarrow h+4$ leaves
the dimer configuration unchanged. Finally, the leading terms in an
effective action incorporating all terms not ruled out by symmetry or
microscopic considerations of the model in $d=2+1$ dimension are
\begin{eqnarray}
{\cal S}=\int d^2 x d\tau &&[(\partial_\tau h)^2+\rho_2(\nabla h)^2\nonumber\\
&+&\lambda\cos(2 \pi h)+\mu \cos(4 \pi h)]\ ,
\end{eqnarray}
where $\tau$ denotes Euclidean time.
This action differs from that in Ref.~\onlinecite{Cantor} in two ways:
(i) a term $(\nabla^2h)^2$ is missing because it plays an
important role only at the RK point where $\rho_2$ vanishes, whereas
we are interested in what happens elsewhere; (ii) the term $\mu
\cos(4 \pi h)$ has been added as its presence is necessary for the
analysis of the transition out of the columnar phase.
To see this, let us minimise ${\cal S}$ for uniform configurations. For
$\mu=0$, the sign of $\lambda$ determines if one has a columnar
($\lambda>0)$ or plaquette state. At the transition point between
the two, $\lambda=0$, implying a continuous degeneracy
($\langle h\rangle$ can take any value $\in [0,4[$), which has so far
not been observed.

For $\mu\neq0$, the fate of this transition depends on the sign of
$\mu$. For $\mu<0$, the minima of the $\lambda$ term coincide with
those of the $\mu$ term, independently of the sign of $\lambda$. This
destroys the continuous degeneracy at $\lambda=0$, and leads to a
simple first order transition between columnar and plaquette
states. By contrast, for $\mu>0$, the $\mu$ and $\lambda$ terms
compete: the minima of the latter coincide with the maxima of the
former. Up to symmetries, one obtains $\langle h\rangle$ changing
continuously between $0$ at $\lambda=-4\mu$ and $1/2$ at
$\lambda=4\mu$ as
\begin{eqnarray}
2\pi\langle h\rangle =\begin{cases}
0\ ,\ \rm{if} \ \lambda<-4\mu, \\
 \arccos(-\lambda/4\mu)\ ,\ \rm{if}\  |\lambda|<4\mu, \\
\pi\ , \ \rm{if}\ \lambda>4\mu\  . 
\end{cases}
\end{eqnarray}
In the region $ |\lambda|<4\mu$ we therefore find a low-symmetry 
mixed state which
continuously crosses over from primarily columnar to plaquette
character. If $-\lambda/4\mu$ reaches 1 before $v/t$ does, the mixed
phase terminates before the RK point; otherwise, it will be terminated
by the RK transition. We note that other 'non-generic' scenarios (e.g.\
for
particularly small values of $\mu$, or for non-monotonic dependence of
$\lambda/\mu$ on $v/t$) are also possible.

\emph{Concluding remarks}: The analysis of low energy spectra offers a
powerful method to investigate spontaneous symmetry breaking, in
particular eliminating the bias that the choice of an order
parameter entails. 
While our system sizes -- up to 30$\times$30
sites with GFMC -- are much bigger than those usually accessible to
ED, they are still subject to the uncertainties of extrapolating
delicate ordering phenomena to the thermodynamic limit. 

However, our analytics support the numerics by showing that 
the phase diagram incorporating an intermediate mixed phase is 
one of two generic possibilities for such RK models, the
other being a direct first order columnar-plaquette transition.  The square
lattice QDM is the first strong candidate for realising this new
scenario.

\emph{Acknowledgements}. D.P. and A.R. acknowledge support from the
Agence Nationale de la Recherche (France).
R.M. and D.P. thank the Kavli Institute for Theoretical Physics 
for hospitality during the last stage of this work.
This research was supported in part by the National Science 
Foundation under Grant No. PHY05-51164.


\end{document}